\documentclass[useAMS,usenatbib]{mn2e}
\usepackage{graphicx}

\newcommand{\Teff}{\mbox{$T_{\mathrm{eff}}$}}

\newcommand{\Lines}[3]{\Ion{#1}{#2}\,#3}
\newcommand{\Ion}[2]{#1{\,\small #2}}

\newcommand{\Mwd}{\mbox{$M_{\mathrm{wd}}$}}
\newcommand{\Msun}{\mbox{$\mathrm{M}_{\odot}$}}
\newcommand{\Rsun}{\mbox{$\mathrm{R}_{\odot}$}}

\title[The first  DBZ white dwarf with a
metal-rich gaseous debris disc]{SDSS\,J084539.17+225728.0: the first  DBZ white dwarf with a
metal-rich gaseous debris disc}

\author[B.T. G\"ansicke et al.]{
B.T. G\"ansicke$^1$,
D. Koester$^2$,
T.R. Marsh$^1$,
A. Rebassa-Mansergas$^1$,
J. Southworth$^1$\\
$^{1}$ Department of Physics, University of Warwick, Coventry CV4 7AL,
UK \\
$^{2}$ Institut f\"ur Theoretische Physik und Astrophysik, University of Kiel,
24098 Kiel, Germany\\
}

\begin{document}

\date{Accepted 2005. Received 2005; in original form 2005}

\pagerange{\pageref{firstpage}--\pageref{lastpage}} \pubyear{2006}

\maketitle

\label{firstpage}

\begin{abstract}
We report the discovery of a third white dwarf hosting a gaseous
debris disc, SDSS\,J084539.17+225728.0. The typical double-peaked
\Lines{Ca}{II}{8498,8542,8662}\,\AA\ emission lines can be
modelled in terms of a Keplerian gas disc with a radial extent from
$\sim0.5$\,\Rsun\ to $\sim1.0$\,\Rsun.  The effective temperature of
SDSS\,0845+2257, $\Teff\simeq18600\pm500$\,K, is comparable to the two
other white dwarfs with gaseous discs, SDSS\,1043+0855 and
SDSS\,1228+1040, and hence substantially hotter than the bulk of white
dwarfs where dusty debris discs were identified through the presence
of infrared excess flux. This may suggest that the conditions to
produce emission lines from debris discs in the optical wavelength
range are only met for a relatively narrow range in $\Teff$.  The
observed asymmetry in the line profiles indicates a substantial
eccentricity in the disc. Two spectra obtained four years apart reveal
a significant change in the shapes and equivalent widths of the line
profiles, implying that the circumstellar disc evolves on relatively
short time scales. In contrast to SDSS\,1043+0855 and SDSS\,1228+1040,
SDSS\,0845+2257 has a helium-dominated atmosphere. We detect
photospheric absorption lines of He, Ca, Mg, and Si in the
Sloan Digital Sky Survey spectrum, and hence classify SDSS\,0845+2257
as DBZ white dwarf. The abundances for the three metals determined
from model atmosphere fits are $\mathrm{Ca/He}\simeq1.3\times10^{-7}$,
$\mathrm{Mg/He}\simeq6.0\times10^{-6}$, and
$\mathrm{Si/He}\simeq8.0\times10^{-6}$. From the non-detection of
H$\alpha$ we derive $\mathrm{H/He}<3\times10^{-5}$, which implies that
the hydrogen-to-metal abundance ratio of the circumstellar material is
$\ga1000$ times lower than in the Sun. This lends strong support to
the hypothesis that the gaseous and dusty debris discs found around
roughly a dozen white dwarfs originate from the disruption of rocky
planetary material.
\end{abstract}


\section{Introduction}
Recent years have seen a surge of interest in the evolution of
extra-solar planetary systems through the late phases in the evolution
of their host stars.  \citet{ignace01-1}, \citet{chuetal01-1}, and
\citet{burleighetal02-1} suggested that Jovian planets around white
dwarfs may be detected either through direct imaging or
spectroscopy. While the survival of planets seems plausible
\citep{villaver+livio07-1}, there is as yet no unambiguous detection of a
planet around a white dwarf (\citealt{burleighetal08-1}, but see also
\citealt{mullallyetal08-1} for a good candidate).

\begin{figure}
\centerline{\includegraphics[angle=0,width=0.75\columnwidth]{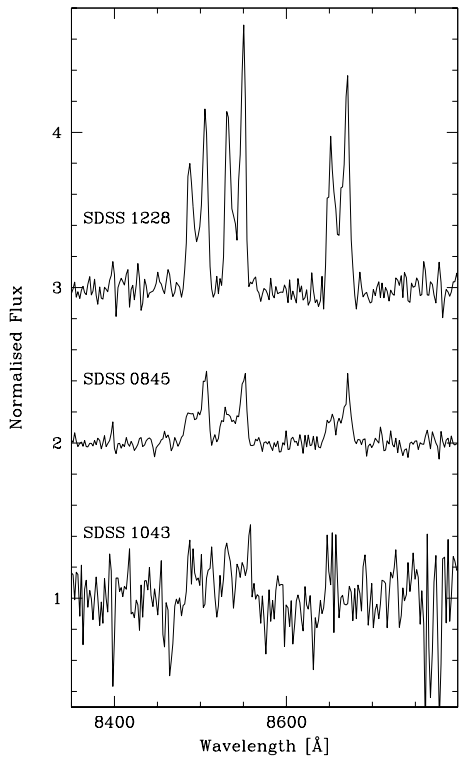}}
\caption{\label{f-caii} Comparison of the
  \Lines{Ca}{II}{8498,8542,8662}\,\AA\ emission lines in the SDSS spectra of
  the three white dwarfs SDSS\,1228+1040 (top,
  \citealt{gaensickeetal06-3}), SDSS\,0845+0855 (middle) and
  SDSS\,1043+0855 (bottom, \citealt{gaensickeetal07-1}). The spectra
  of SDSS\,0845+0855 and SDSS\,1228+1040 are offset by one and two
  units, respectively. }
\end{figure}

However, as part of a search for cool companions to white dwarfs, an
infrared excess was discovered around the cool white dwarf G29--38
\citep{zuckerman+becklin87-1}. HST observations of G29--38 revealed a
large photospheric abundance of metals, implying that the star is
accreting at a relatively high rate \citep{koesteretal97-1}.  Deep
imaging and asteroseismological studies of G29--38 ruled out a brown
dwarf companion \citep{grahametal90-1, kleinmanetal94-1,
  kuchneretal98-1} and led to the hypothesis of a cool dust cloud
around the white dwarf. The presence of dust near G29--38 has been
verified by infrared observations with the \textit{Spitzer Space
  Telescope} \citep{reachetal05-1}. Recent infrared surveys have
boosted the number of cool white dwarfs harbouring dust discs to
$\sim10$ (e.g. \citealt{becklinetal05-1, kilicetal05-1, kilicetal06-1,
  kilic+redfield07-1, vonhippeletal07-1, juraetal07-1,
  farihietal08-1}). These dust discs originate from the tidal
disruption of either comets \citep{debesetal02-1} or asteroids
\citep{jura03-1}. Asteroids appear to be more likely as they can
explain the large metal abundances in the material accreted by the
white dwarfs from the dusty environment.

Hydrogen-rich white dwarfs with radiative atmospheres
($\Teff\ga11500$\,K) have short diffusion time scales for heavy
elements, and hence their photospheric abundances should closely
reflect the chemical abundances of the circumstellar debris
discs from which they are accreting. \citet{zuckermanetal07-1}
suggested that such systems offer potential insight into the chemical
composition of extra-solar planetary systems, once more detailed
calculations of diffusion time scales for a wide range of elements and
atmospheric parameters are available.

We have recently identified two hydrogen-dominated (DA) white dwarfs,
SDSS\,J122859.93+104032.9 and SDSS\,J104341.53+085558.2 (henceforth
SDSS\,1228+1040 and SDSS\,1043+0855), that exhibit double-peaked
emission lines in the $I$-band \Ion{Ca}{II} triplet
\citep{gaensickeetal06-3, gaensickeetal07-1}. From the morphology of
the \Ion{Ca}{II} lines profiles we inferred the presence of gaseous
circumstellar discs close ($\sim R_\odot$) to the white dwarfs.  Like
the majority of the white dwarfs with dusty debris discs, both
SDSS\,1228+1040 and SDSS\,1043+0855 have hydrogen-rich metal-polluted (DAZ)
atmospheres, indicating that they are accreting from the circumstellar
material. The absence of photospheric helium absorption lines, as well
as of Balmer emission lines from the gaseous discs, strongly suggest
that circumstellar material around SDSS\,1228+1040 and SDSS\,1043+0855
is depleted in volatile elements. A dusty extension to the gaseous
disc in SDSS\,1228+1040 has been detected with \textit{Spitzer}
\citep{brinkworthetal08-1}.

Here we present a third white dwarf with a metal-rich gaseous debris
disc, the helium-dominated (DB) SDSS\,J084539.17+225728.0 (henceforth
SDSS\,0845+2257).

\begin{figure*}
\includegraphics[angle=-90,width=12.8cm]{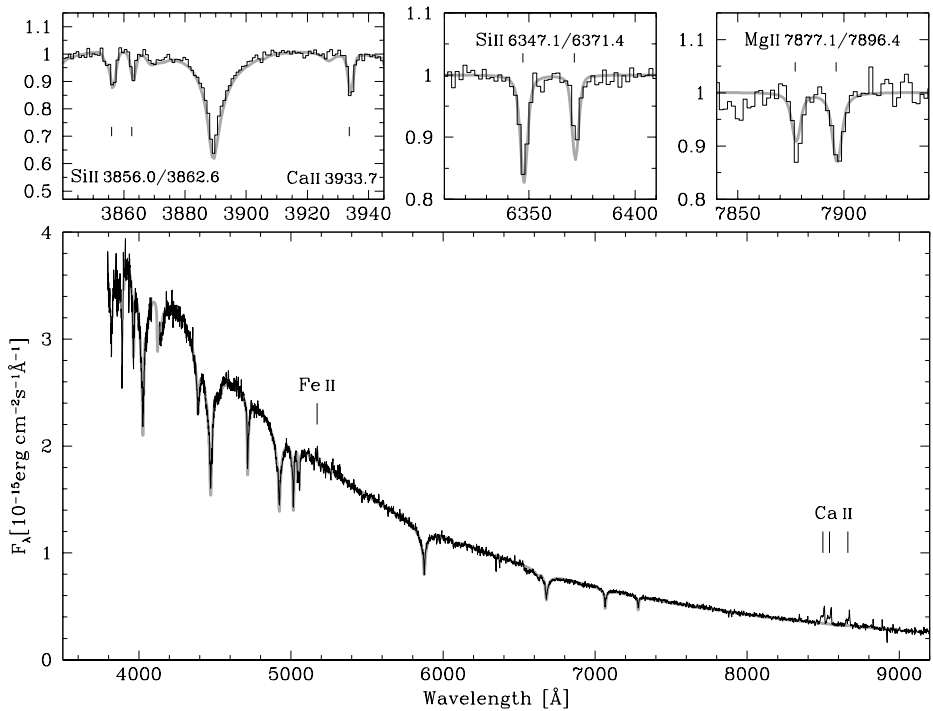}
\caption{\label{f-opt_fit} Model atmosphere fit to the spectrum of
  SDSS\,0845+2257. Main panel: best-fit to the SDSS spectrum with
  $\Teff=18621$\,K and $\log g=8.175$. The positions of the
  \Ion{Ca}{II} and \Ion{Fe}{II} emission lines are indicated. Top
  panels: close-ups of the average WHT spectrum regions featuring
  photospheric metal absorption lines Ca, Mg, and Si. The model shown
  has been computed for the following abundances:
  $\mathrm{Ca/He}\simeq1.3\times10^{-7}$,
  $\mathrm{Mg/He}\simeq6.0\times10^{-6}$, and
  $\mathrm{Si/He}\simeq8.0\times10^{-6}$. The non-detection of H$\alpha$
  absorption implies $\mathrm{H/He}<3\times10^{-5}$. }
\end{figure*}

\section{SDSS and WHT observations of SDSS\,0845+2257}
We selected 15116 unique spectroscopic objects from Data
Release\,6 of the Sloan Digital Sky Survey (SDSS DR\,6,
\citealt{adelman-mccarthyetal08-1}) that were classified by the
SDSS spectroscopic reduction pipeline as stars, and are bluer
than the main sequence ($u-g<0.8$) with relatively low uncertainties
in their $i$-band photometry ($\sigma_i<0.1$\,mag). The spectra of
these objects were subjected to an automatic measurement of the
equivalent width of the \Lines{Ca}{II}{8498,8542,8662}\,\AA\
triplet (see \citealt{gaensickeetal07-1} for details). This search
resulted in a list of 484 candidates exhibiting a $3\sigma$ excess in
the \Ion{Ca}{II} triplet over the neighbouring continuum, which were
then visually inspected. The vast majority of the spectra flagged by
our program are affected by poor night sky line
subtraction\footnote{Sky background subtraction in fibre
    spectroscopy, such as carried out by SDSS, is notoriously more
    difficult than in slit spectroscopy as (a) the fibres typically
    have a larger on-sky area ($3\arcsec$ in the case of SDSS) than a
    slit, resulting in a larger contamination by sky background, and
    (b) the sky subtraction is done via dedicated sky fibres that are
    placed some distance away from the target fibre, which can result
    in different optical paths, spectroscopic responses, etc. In
    addition, any small offset in wavelength between the sky spectrum
    and object spectrum will result in large residuals near strong sky
    lines.}. Among the 484 candidates, we identified 82
cataclysmic variables (CVs) exhibiting Ca\,II emission, most of which
were previously discovered \citep[][and references
  therein]{szkodyetal07-2}. The spectra of all those CVs
  contain strong and broad Balmer emission lines, and in most cases
  noticeable He emission lines as well.

Only three apparently single white dwarfs with unambiguous emission in
all three components of the \Ion{Ca}{II} triplet were found
(Fig.\,\ref{f-caii}): the two previously known DA white dwarfs with
gaseous debris discs, SDSS\,1228+1040 and SDSS\,1043+0855, and the
first DB white dwarf displaying \Ion{Ca}{II} emission, SDSS\,0845+2257
(Fig.\,\ref{f-opt_fit}). SDSS\,0845+2257 has been previously
identified as a UV-excess source (PG\,0842+231, Ton 345) and
classified as an sdO subdwarf \citep{greenetal86-1}. The absence of
close companion to these three white dwarfs is suggested by the
complete lack of either Balmer or He emission lines in their optical
spectra. The single white dwarf nature of SDSS\,1228+1040 was
corroborated by extensive time-resolved spectroscopy and photometry,
which ruled out the presence of short-period radial velocity
variations and 
brightness modulations \citep{gaensickeetal06-3}. By analogy to
SDSS\,1228+1040, we infer that SDSS\,0845+2257 is also a single white
dwarf harbouring a metal-rich gaseous accretion disc fed from
circumstellar debris.

The double-peaked \Ion{Ca}{II} profiles in SDSS\,0845+2257 display a strong
asymmetry (Fig.\,\ref{f-caii}), suggesting a stronger ellipticity
and/or azimuthal brightness variation of the gaseous ring than in the
other two systems. Closer inspection of the SDSS spectrum revealed
also weak emission of Fe\,II 5169\,\AA, as well as narrow, weak
absorption lines of Ca, Si, and Mg originating in the helium-dominated
atmosphere of the white dwarf.

\begin{figure*}
\includegraphics[angle=-90,width=13.5cm]{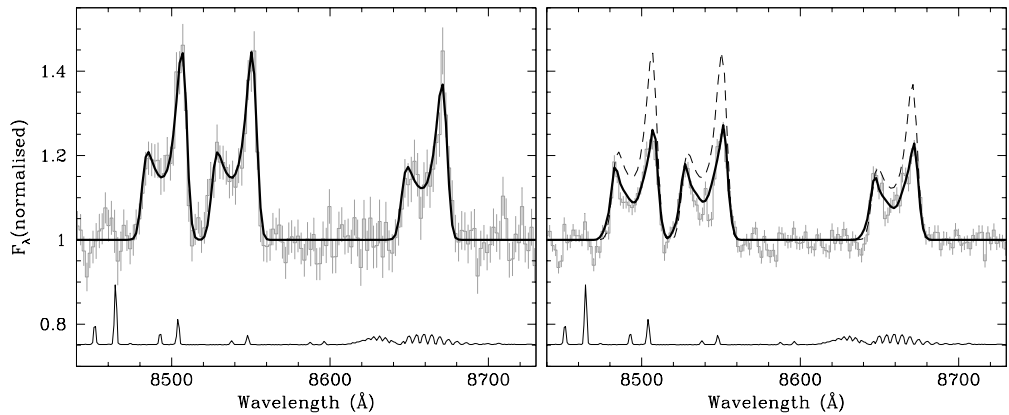}
\caption{\label{f-caii_time} Comparison of the
  \Lines{Ca}{II}{8498,8542,8662}\,\AA\ emission lines of
  SDSS\,0845+2257 in the SDSS spectrum obtained in December 2004
  (left) and in our WHT spectrum from January 2008 (right). The data
  are shown as gray histogram with error bars. Besides a clear change
  in the morphology of the line profiles the equivalent width
  decreased from $20\pm1$\,\AA\ in 2004 to $13\pm0.5$\,\AA\ in
  2008. Superimposed as thick black lines are the best-fit dynamical
  disc models for the two epochs (Sect.\,\ref{s-ring}). The model for
  the 2004 SDSS spectrum is repeated in the plot of the WHT as thin
  dashed line to illustrate the changes in the line shapes and
  strengths. Finally, the thin solid lines at the bottom of the panels
  indicate the positions of strong night sky lines.}
\end{figure*}

Intermediate resolution spectroscopy of SDSS\,0845+2257 was obtained
on January 2, 2008 using the double-arm spectrograph ISIS on the
William Herschel Telescope. In the blue arm, we used the R600B
grating and a 4k$\times$2k pixel EEV detector, and in the red arm the
R316R grating and a low-fringing 4k$\times$2k pixel REDPLUS
detector. This setup provided a wavelength coverage of
$\simeq3600-5100$\,\AA\ at a resolution of $\simeq1.8$\,\AA\ and
$\simeq6100-8800$\,\AA\ at a resolution of $\simeq3.3$\,\AA. Two pairs
of blue/red spectra with individual exposure times of 15\,min were
obtained under average atmospheric conditions, i.e. thin cirrus and
$\sim1.2\arcsec$ seeing. The data
were reduced following the methods described in
\citet{southworthetal07-2}. The WHT spectra confirmed the presence of
photospheric Ca, Si, and Mg absorption lines, and, more importantly,
reveal a significant change in the shape and strength of the
\Ion{Ca}{II} triplet with respect to the SDSS
spectrum. Fig.\,\ref{f-caii_time} compares the SDSS and the WHT
spectra, and also indicates the wavelengths of strong night sky
lines. It is clear that the observed change in the \Ion{Ca}{II} line
profiles is intrinsic to SDSS\,0845+2257, and not caused by
differences in the sky subtraction between the SDSS and WHT spectra.

\section{Model atmosphere analysis}
We fitted the SDSS spectrum of SDSS\,0845+2257 following the methods
outlined in \citet{vossetal07-1} using the synthetic spectra
calculated with the model atmosphere code described by
\citet{finleyetal97-1, koester+wolff00-1,
  koesteretal05-1}. Using pure He atmosphere models, the best-fit is
achieved for $\Teff=18621\pm60$\,K and $\log g=8.175\pm0.023$.

In a second step, we have investigated the abundances of H, Ca, Mg,
and Si. From the absence of H$\alpha$ absorption, we determine
$\mathrm{H/He}<3\times10^{-5}$. A model calculated with
$\mathrm{Ca/He}\simeq1.3\times10^{-7}$,
$\mathrm{Mg/He}\simeq6.0\times10^{-6}$, and
$\mathrm{Si/He}\simeq8.0\times10^{-6}$ provides an acceptable fit of
the observed photospheric absorption, with an estimated uncertainty of
a factor two in the abundances. A more detailed analysis will have to
await data of higher spectral resolution and signal-to-noise ratio,
and if possible extension of the spectroscopic observations into the
ultraviolet which contains thousands of lines of a large number of
chemical elements.

Adopting a hydrogen abundance of $\mathrm{H/He}=10^{-5}$ results only
in modest changes in the best-fit effective temperature and surface
gravity, $\Teff=18525\pm60$\,K and $\log g=8.281\pm0.022$. The quoted
errors are of purely statistical nature, and should be regarded as a
strict lower limit. Based on our experience, we suggest that realistic
errors are 500\,K for $\Teff$ and 0.2\,dex for $\log g$. Using Wood's
(\citeyear{wood95-1}) mass-radius relation and cooling models, we find
a mass of $\simeq0.7M_\odot$ and a cooling age of $\simeq150$\,Myr.

\section{The structure of the gaseous ring}
\label{s-ring}

The double-peaked Ca\,II emission lines changed significantly both in
morphology and equivalent width  between the SDSS spectrum taken in
December 2004, and our WHT spectrum taken in January 2008
(Fig.~\ref{f-caii_time}). These changes clearly demonstrate that the
circumstellar material is undergoing some kind of evolution on time
scales as short as years. 

We have modelled both sets of observed Ca\,II emission line profiles
(Fig.\,\ref{f-caii_time}) following the prescription outlined in
\citet{gaensickeetal06-3}. In brief, the disc is represented by a set
of co-aligned elliptical orbits of constant eccentricity. The
emissivity varies as a power law in radius and varies azimuthally as a
sinusoid with maximum and minimum flux along the semi-major axis. The
only difference from \citet{gaensickeetal06-3} is that we automated
the optimisation of the model to minimise $\chi^2$. The SDSS spectrum
can be modelled by emission from a disc with an inner and outer radius
of $\approx0.3$\,\Rsun\ and $\approx0.9$\,\Rsun, with
an eccentricity of $\approx0.4$. The WHT spectrum is best-fit
by a disc with an inner radius of $\approx0.5$\,\Rsun\ and
$\approx0.8$\,\Rsun, with an eccentricity of
$\approx0.2$. Our model suffers from degeneracies between
different parameters, and consequently these numbers should be
considered as guidance, rather than exact results. However, the
following conclusions appear to be firm. SDSS\,0845+2257 hosts an
eccentric circumstellar disc of metal-rich gas with Ca\,II emission
originating over a radial extent of $\sim0.5$\,\Rsun\ to
$\sim1.0$\,\Rsun. While the radial extension of the Ca\,II gas is
similar to that found in SDSS\,1228+1040 \citep{gaensickeetal06-3},
the eccentricity of the disc in SDSS\,0845+2257 is substantially
larger. We have initiated a spectroscopic monitoring of
SDSS\,1228+1040 and SDSS\,0845+2257 that will hopefully provide more
insight into the dynamical evolution of their circumstellar discs.

\section{The nature of the circumstellar material}
The origin of metals in the atmospheres of cool white dwarfs has been
subject to long debate, with accretion from the interstellar medium
\citep[e.g.][]{dupuisetal93-2} having been the most widely accepted
scenario until recently. The discovery of circumstellar dust discs
around a number of white dwarfs with metal-polluted atmospheres
\citep[e.g.][]{becklinetal05-1, kilicetal05-1, farihietal08-1} led to
the suggestion that accretion of comets \citep{debesetal02-1} or
asteroids \citep{jura03-1} may contribute to the pollution of white
dwarfs with heavy elements. 

In the case of DAZ white dwarfs, which have hydrogen-dominated
atmospheres, the relative abundances of hydrogen and metals within the
accreted material are difficult to gauge. In contrast to this,
metal-polluted white dwarfs with helium-rich atmospheres allow much
tighter constraints on the hydrogen abundance of the accreted
material. \citet{dupuisetal93-2} and \cite{friedrichetal99-1} showed
for the DBZ white dwarfs GD\,40 and HS\,2253+8023 that whereas the
metal-to-metal abundance ratios were consistent with accretion from
the ISM, the hydrogen-to-metal abundance ratio has to be $10^4-10^6$
lower than typical ISM abundances to be consistent with the observed
spectra of the two DBZ white dwarfs. In a later paper,
\cite{friedrichetal04-1} ruled out that a weak magnetic field acting
as a propeller might suppress the accretion of hydrogen in
GD\,40. Recently, \cite{juraetal07-1} detected infrared excess around
GD\,40 which strongly supports the hypothesis that the metals in the
atmosphere of the white dwarf are related to ongoing accretion from a
disc of hydrogen-depleted dust. \cite{jura06-1} took the argument of
accretion of asteroid debris in GD\,40 a step further, suggesting that
the apparent carbon-to-iron deficiency in the atmosphere of the star
is similar to the abundance pattern of chondritic asteroids.

Besides GD\,40, SDSS\,0845+2257 is only the second bona-fide DBZ white
dwarf in which a circumstellar disc has been detected. Our current
observations constrain the metal-to-hydrogen abundance ratio of the
accreted material to be $10^3-10^4$ times larger than observed in the
Sun, which strongly supports \citeauthor{jura03-1}'s
(\citeyear{jura03-1}) scenario of metal pollution by ongoing accretion
from tidally disrupted asteroids. 

While DB white dwarfs allow tight constraints on the hydrogen fraction
in the accreted material, the sedimentation times for metals are many
orders of magnitude larger than in DA white dwarfs
(\citealt{paquetteetal86-1, koester+wilken06-1}; e.g. $\sim10^5$\,yr
for SDSS\,0845+2257 vs. days in SDSS\,1228+1040), and may differ by
factors of a few for different chemical species. In order to reach an
accretion/diffusion steady state, the white dwarf has to accrete at a
constant rate for 2--3\,times the diffusion time scale, i.e. close to
$10^6$\,yr in the case of SDSS\,0845+2257. It seems unlikely that an
asteroid debris disc can sustain a constant mass loss rate for such a
long period of time. In the case of time-dependent accretion rates,
the observed metal-to-metal abundance ratios will be
variable. \citet{dupuisetal93-2} presented a time-dependent model of
accretion onto white dwarfs, adopting a variation in accretion rate by
a factor $10^5$, and found that observed Mg/Ca, Si/Ca, and Fe/Ca
metal-to-metal ratios could differ by up to a factor 100, depending on
the timing of the observations with respect to the change in accretion
rate. We conclude that, given the long diffusion time scales, the
metal-to-metal abundance ratios seen in a DBZ atmosphere are unlikely
to be an immediate proxy for the chemical abundances of the
circumstellar material.

\section{Conclusions}
We have identified SDSS\,0845+2257 as the third white dwarf known to
host a metal-rich gaseous circumstellar disc. SDSS\,0845+2257 has a
similar temperature as SDSS\,1043+0855 and SDSS\,1228+1040, and all
three are substantially hotter ($\Teff\sim20\,000$\,K) compared to the
majority of the white dwarfs with dusty discs identified in surveys
for infrared excess fluxes. The total roster of white dwarfs with
gaseous and/or dusty debris discs stands at 13, of which $\sim25$\%
were identified through the presence of \Ion{Ca}{II} emission lines in
their SDSS spectra. Taking the masses of the three white dwarfs with
gaseous debris discs at face value, we find
$<\Mwd>=0.73\pm0.07\,\Msun$, which is marginally higher than the mean
mass of field white dwarfs \citep{liebertetal05-1}. A speculative line
of thought is that the potentially higher occurence of debris discs
around more massive white dwarfs is related to the higher frequency of
debris discs around A-stars compared to Sun-like stars
\citep{suetal96-1, trillingetal08-1}. 

In contrast to SDSS\,1228+1040 and SDSS\,1043+0857, SDSS\,0845+2257
has a helium-dominated atmosphere and the detection of Ca, Mg, and Si
absorption lines qualifies it as a DBZ white dwarf. The presence of a
circumstellar disc strongly suggests that ongoing accretion is the
origin of the observed photospheric metals. The non-detection of
H$\alpha$ absorption in the SDSS spectrum of SDSS\,0845+2257 implies
that the hydrogen-to-metal ratio in the circumstellar material is at
least a factor 1000 smaller than the solar value
\citep{dupuisetal93-2, friedrichetal99-1}. The most likely
origin of the metal-rich material is the planetary debris, such as the
tidal disruption of a rocky asteroid \citep{jura03-1}.

Modelling the broad double-peaked \Ion{Ca}{II} emission lines observed
in SDSS\,0845+2257 with a Keplerian gas disc confines the region of
the \Ion{Ca}{II} emission to a radial extension of
$\sim0.5$\,\Rsun\ to $\sim1.0$\,\Rsun. The line profiles display a
large asymmetry, which we interpret as a significant asymmetry in the
gaseous disc. The line profile shapes and equivalent widths of the
\Ion{Ca}{II} emission lines varied substantially in between two
observations taken $\sim4$ years apart. Following the detection of
variability in the equivalent widths of the photospheric \Ion{Ca}{H,K}
absorption lines lines seen in G29--38 \citep{vonhippel+thompson07-1},
this is the second piece of evidence indicating that the structure of
circumstellar debris around white dwarfs undergo changes on relatively
short time scales. \citet{jura08-1} suggested that rather than a
single tidal disruption event, white dwarfs with remnants of planetary
systems may experience repeated accretion of small asteroids. Such
recurrent events may potentially be related to variability such as
observed in SDSS\,0845+2257. Whatever the cause of the changes in the
\Ion{Ca}{II} emission lines, long-term monitoring of SDSS\,0845+2257
is likely to offer direct dynamical insight into the evolution of
debris discs around white dwarfs.

\vspace*{-3ex}
\section*{Acknowledgements}
Funding for the Sloan Digital Sky Survey (SDSS) and SDSS-II has been
provided by the Alfred P. Sloan Foundation, the Participating
Institutions, the National Science Foundation, the U.S. Department of
Energy, the National Aeronautics and Space Administration, the
Japanese Monbukagakusho, and the Max Planck Society, and the Higher
Education Funding Council for England. The SDSS Web site is
http://www.sdss.org/. We thank the referee, Mukremin Kilic, for his
quick and constructive report.

\bsp

\label{lastpage}

\end{document}